\documentclass[twocolumn,superscriptaddress,floatfix,showpacs,aps]{revtex4}
\usepackage{amsmath,amssymb,epsfig,color}
\usepackage{graphicx}
\usepackage{dcolumn}
\usepackage{bm}
\usepackage{color}

\begin{document}

\title{Multipeak quasielastic light scattering and high-frequency electronic excitations in honeycomb Li$_2$RuO$_3$.}
\pacs{}
\author{Yuri S. Ponosov}
\affiliation{M.N. Mikheev Institute of Metal Physics UB RAS, 620137, S. Kovalevskaya str. 18, Ekaterinburg, Russia}
\affiliation{Ural Federal University, Mira St. 19, 620002 Ekaterinburg, Russia}

\author{Sergey V. Streltsov}
\affiliation{M.N. Mikheev Institute of Metal Physics UB RAS, 620137, S. Kovalevskaya str. 18, Ekaterinburg, Russia}
\affiliation{Ural Federal University, Mira St. 19, 620002 Ekaterinburg, Russia}

\begin{abstract}
We measured the temperature dependence of low-frequency Raman spectra in Li$_2$RuO$_3$, and  observed multipeak quasielastic scattering in the Ru honeycomb polarizations  below and above the magnetostructural transition temperature. We attribute this scattering to the fluctuations of the energy density in the spin system.  High-frequency electronic light scattering was observed at 2150 $cm^{-1}$. Its intensity increased significantly below the transition temperature, confirming substantial modification of electronic structure due to removal of degeneracy in $t_{2g}$-manifold of Ru$^{4+}$ ions.
\end{abstract}    

\date{\today}

\maketitle

\section{Introduction}
The cluster Mott insulators are one of the most interesting group of materials nowadays. They differ from conventional Mott materials because one needs to consider not a single ion as a correlated unit, but several ones. This gives additional freedom to such systems and they may have rather unusual physical properties, e.g., one may observe suppression of a magnetic response because of condensation of part of the spins into singlets~\cite{Sheckelton2012,Chen2018a,Nikolaev2020}, orbital selective behaviour~\cite{Streltsov2014,Streltsov2016b}, stabilization of the charge ordered state due to dimerization~\cite{Gapontsev2016} or formation of a spin-liquid ground state~\cite{Haraguchi2015,Akbari-Sharbaf2018}. There are clusters in these materials formed by transition metals, where electrons behave as practically delocalized, but still correlated, whereas hopping between such clusters can be rather weak. While in some situations these compounds can be described by methods typically used for isolated molecular systems, the most interesting behaviour they demonstrate close to a localized-itinerant crossover, i.e. close to the Mott transition, where both experimental and theoretical investigation of such materials becomes rather complicated.

Li$_2$RuO$_3$ represents one of the examples of such systems. This material has a layered structure with Ru ions forming a honeycomb lattice. There occurs at T$_C\sim 540$ K a phase transition with the formation of Ru dimers~\cite{Miura}, with Ru-Ru distance in the dimers $\sim$2.57 \AA~(shorter than in Ru metal - 2.65 \AA\cite{Streltsov-UFN}). This structural transition is accompanied by a strong decrease of the magnetization suggesting stabilization of a spin singlet state with $S_{tot}=0$ out of $S=1$ (Ru ions are 4+ with $t_{2g}^4$ electronic configuration). Below T$_C$ dimers form a herring-bone pattern and whole structure can be described by the P2$_1$/m space group. The microscopic reason for the dimerization seems to be a strong direct $d-d$ hopping in the common edge geometry (two RuO$_6$ octahedra share their edges). {\it Ab initio} calculations in general support this picture~\cite{Zlata,Mazin}, although it seems that in reality also the hopping via oxygens is not negligible. The X-ray diffraction measures an average structure (characterized by the C2/m structure\cite{Miura}) without any dimers above 540 K, but the pair distribution function analysis clearly shows that they are extremely stable and exist even at 900 K~\cite{Mazin}.

In order to explain the origin of this transition, different scenarios such as the transition from a highly correlated metal to a molecular-orbital insulator accompanied by bond-dimer formation~\cite{Miura} and the formation of nonmagnetic dimerized superstructure by magneto-elastic mechanism~\cite{Jackeli} have been proposed. In the last study, it was found that the observed superstructure of antiferromagnetic bonds is energetically more favorable than long short chains. It has been proposed that once one of the dimers is formed, the contraction induces a shift of neighboring Li ion, resulting in a dimerization in next Ru-Ru pairs.  This mechanism has been called cooperative ``dimer Jahn-Teller'' effect. Also, the weak inter-dimer coupling may, at least, partially lift the orientational degeneracy through order out of disorder by triplet fluctuations~\cite{Jackeli,Iv}. However, this theory assumes that $d$ electrons in Li$_2$RuO$_3$ are strongly correlated and can be described by the Kugel-Khomskii type spin-orbital Hamiltonian~\cite{Kug}, while in fact there is a strong bond-antibonding splitting in Ru $4d$ and molecular-orbital treatment seems to be rather natural approach in this case~\cite{Mazin}.

Thus, the origin of the observed coupling between the lattice and magnetic features in Li$_2$RuO$_3$ is not completely understood. Authors of~\cite{Mech} observe that for Li$_2$RuO$_3$ the transition might occur in two steps with a first-order structural transition occurring first near 570 K which then drives the magnetic Ru-Ru dimerization transition near 540 K. Our previous Raman scattering study of  Li$_2$RuO$_3$~\cite{Ram19} has revealed the existence of anomalies in phonon self-energies near T$_C$. The observed anomalies suggest their connection either with possible dynamic disordering of Ru atomic positions or with spin-phonon interaction. We noted that the phonon broadening map well the  magnetic susceptibility curve, indicating that the phase transition leaves a fingerprint in the phonon dynamics of  Li$_2$RuO$_3$. In addition, there was a significant increase in the electronic background, on which phonon lines are superimposed, with increasing temperature. 

Raman spectroscopy has proven to be an extremely powerful technique to probe magnetic excitations and spin-lattice interactions in a low-dimensional spin system with unprecedented precision~\cite{Lem}. In low-dimensional spin systems quasielastic light scattering is a rather general feature coupled to critical fluctuations of the system.  The main motivation of this work is using Raman spectroscopy technique to observe in details such critical scattering of light, clarify the  electronic structure of Li$_2$RuO$_3$ and determine the relevant energy scales.

\section{Experimental details}
Raman experiments were carried out on freshly cleaved surfaces of compacted disks of Li$_2$RuO$_3$  polycrystalline powder. We used the same samples on which experiments~\cite{Zlata,Arapova,Ram19} were previously performed. These samples of Li$_2$RuO$_3$ were synthesized using a solid-state reaction method as described in ~\cite{Zlata,Arapova}. Measurements of X-ray diffraction and magnetization confirmed the high quality of our samples showing clear signs of the phase transition at around 560 K.

Polarized Raman measurements  in the temperature range of 300 to 840 K were performed in backscattering geometry using RM1000 Renishaw microspectrometer equipped with 532 nm solid-state laser  and 633 helium-neon laser. Respective Linkam stage was used for temperature variation. Most of the measurements were done on hexagonal microcrystals, that is, on an ab (XY) plane, and the long axis is assumed to be the axis b. Polarization measurements on many such crystals gave repetitive spectra for the used in-plane and out-of-plane polarizations. Measurements on thin rectangular crystals (YZ plane) confirmed measurements on hexagonal-shaped crystals. The spectra obtained in ~\cite{Ram19} allow us to state that the experiment was carried out on untwinned crystals, as evidenced by the polarization dependences of just the $B_g$ phonon lines.  The laser beam was focused (~5 $\mu$m in diameter) on microcrystals of hexagonal shape up to 30  $\mu$m in size (XY plane) or on thin rectangular crystals (XZ or YZ plane). Very low power (up to 0.1 mW) was used to avoid local heating of the crystals. A pair of notch filters with cut-off at 60 $cm^{-1}$ was used to reject light from the 633 nm laser line. In order to get as close to the zero frequency as possible, with 532 nm excitation we used a set of three volume Bragg gratings (VBG) to analyze the scattered light. This made it possible to reach frequencies of 10 $cm^{-1}$ and to get an access the anti-stokes spectrum. The resolution of our Raman spectrometer is estimated to be ~ 2-3 $cm^{-1}$.
\begin{figure}[b]
\includegraphics[width=0.45\textwidth]{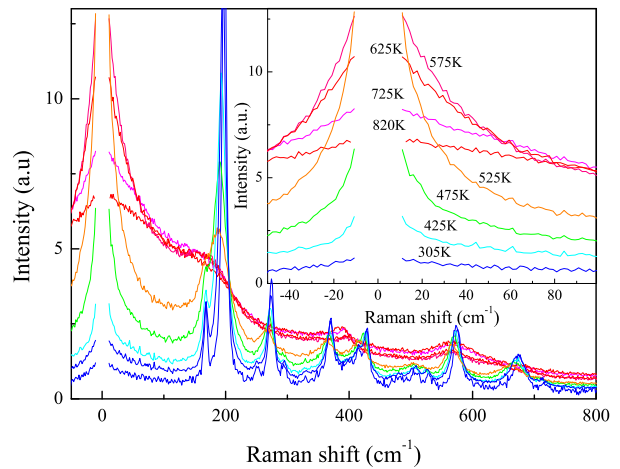}
\caption{\label{asmeash} Experimental Raman spectra of Li$_2$RuO$_3$  measured at different temperatures in the YY polarization geometry. Inset shows low-frequency range in more details. Excitation-532 nm. }
\end{figure} 

\section{Results and discussions}
\subsection{Quasielastic Raman spectra of Li$_2$RuO$_3$}

Fig.~\ref{asmeash} presents the Raman spectra of Li$_2$RuO$_3$ measured at different temperatures in phonon region. Phonon features in these spectra were discussed in our previous article~\cite{Ram19}, where attention was drawn to the unusually strong growth of the background in the low-frequency region. This quasielastic scattering extends to 800 $cm^{-1}$ and its growth with temperature increase is observed mainly in the in-plane YY, XX ($A_g$ symmetry) and XY polarizations ($B_g$ symmetry) of incident and scattered light~\cite{Ram19}. Intensity of  quasielastic scattering at room temperature is somewhat lower ($\sim$30-40\%) in the XY  and  the out-of-plane ZZ ($A_g$ symmetry channel) geometry   than in XX and YY polarisations. An increase in the intensity of in-plane quasielastic scattering is observed at all frequencies already before the structural transition and  reaches low-frequency maximum at $T\sim$525 K (Fig.~\ref{asmeash}), then its intensity in the region below 100-150 $cm^{-1}$ decreases. 

\begin{figure}[b]
\includegraphics[width=0.45\textwidth]{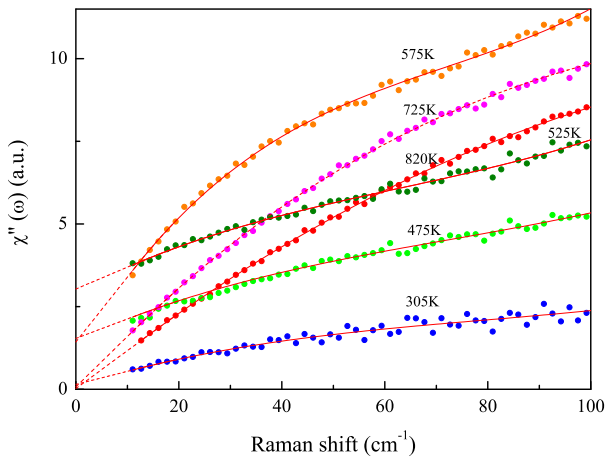}
\caption{\label{RSnew} Low-frequency Stokes Raman response $\chi''(\omega)$ of Li$_2$RuO$_3$ measured in YY polarization geometry at different temperatures. Solid lines are polynomial fits to data, dashed lines indicate possible extrapolations to $\omega$=0.}
\end{figure}

The Raman response $\chi''(\omega)=I(\omega)/[n(\omega)+1]$, presented in Fig.~\ref{RSnew} was obtained from the raw Raman spectra $I(\omega$) (Fig.~\ref{asmeash}), where $n(\omega)$ is the Bose factor. As one can see, the  $\chi''(\omega)$ frequency  behavior shows dramatic changes at $T \geq$ 525K. There is a significant increase of the response measured at 575 K for  $\omega \geq$ 50 $cm^{-1}$, along with a significant change in the frequency behavior below 50 $cm^{-1}$. This suggests a nonzero response at $\omega$ = 0. Although our measurements were performed at $\omega\geq$ 10 $cm^{-1}$, we tried to roughly extrapolate the measured response to $\omega=0$ frequency. Such a procedure is not entirely justified, but it clearly shows presence of a central component of the Raman response, which substantially depends on temperature. Of course, the detailed line shape and its width cannot be determined without using low-frequency measurements, for example, using Brillouin light scattering.

\begin{figure}[b]
\includegraphics[width=0.5\textwidth]{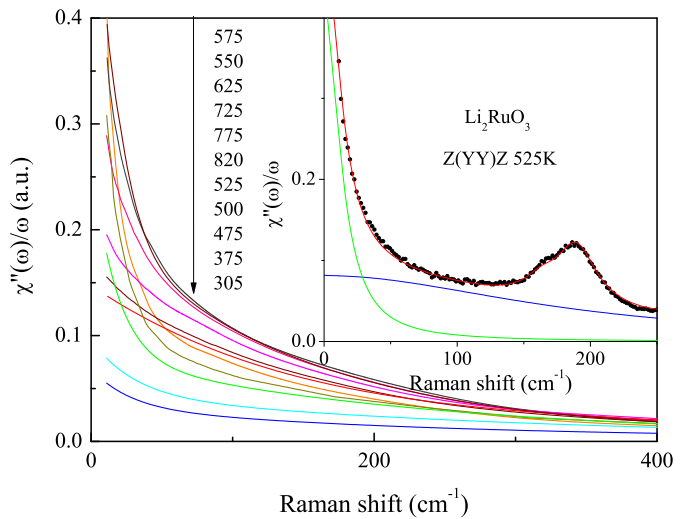}
\caption{\label{Dyn}  Raman conductivities in Li$_2$RuO$_3$ at different temperatures (figures). Inset demonstrates results of fitting of dynamical susceptibility $\chi^{dyn}_{A_{1g}\|}$ at 525 K by two Lorenzians shown by green and blue lines. Two phonon peaks at $\sim$180 cm$^{-1}$ were subtracted before  fitting.}
\end{figure}

Then after subtracting the phonon peaks we defined the Raman conductivity or dynamic susceptibility $\chi''(\omega)/\omega$. The frequency dependences of $\chi''(\omega)/\omega$ at different temperatures are shown in Fig.~\ref{Dyn}. One may see that the YY Raman conductivity features a pronounced peak at $\omega\rightarrow0$. Similar curves were obtained for the XY polarization geometry, i.e. $B_g$ symmetry. Note, that amplitude of $\chi''(\omega)/\omega$ varies strongly with temperature for both polarization geometries, but practically stays constant for the ZZ geometry~\cite{Ram19}. Therefore, further our attention will be focused on the susceptibility in the honeycomb plane. The temperature dependence and the distinctive honeycomb plane A$_g$+$B_g$ symmetry of this low frequency quasielastic peak clearly links it to dynamic fluctuations corresponding to some order in honeycomb plane. We used Lorentzian relaxational forms to fit the Raman conductivity spectra in whole frequency range down to zero as shown in inset of Fig.~\ref{Dyn}. The obtained $\chi''(\omega)/\omega$ can not be fitted by one peak, but an excellent fit by two Lorentzians was obtained for all temperatures having very different widths, but both centered at zero frequency. Since the subtraction of phonon peaks can be ambiguous, especially at high temperatures, full-spectrum fittings were also used to confirm the stability of the obtained Raman conductivity parameters (inset in Fig.~\ref{Dyn}). Good agreement was obtained by both types of fittings. As shown earlier, in addition to these two peaks, there is  possibly a much narrower central peak near the transition.  Extrapolating the response to zero frequencies (Fig.~\ref{Dyn}) implies that the central peak intensity is an order of magnitude higher than the intensity measured at 10 $cm^{-1}$, and its half-width is $\sim$ 1 wavenumber. Therefore, neglecting this contribution when adjusting the dynamic susceptibility curves we do not strongly affect the results.

\begin{figure}[t]
\includegraphics[width=0.45\textwidth]{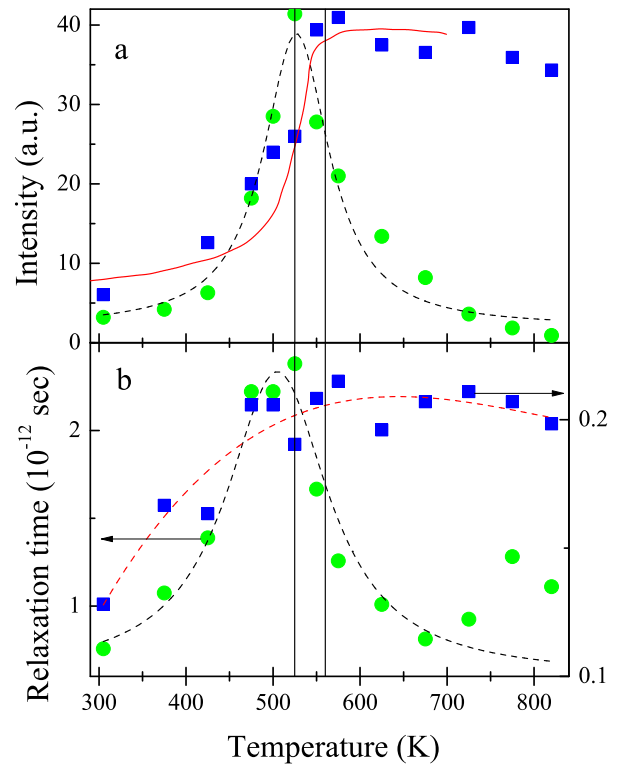}
\caption{\label{peaks}  The temperature dependence of the peak intensity (a) and the relaxation time (b) of the first (narrow Lorenzian in inset of Fig.~\ref{Dyn}; circles here)  and second (wide Lorenzian in inset of Fig.~\ref{Dyn}; squares here) relaxation modes. Magnetic susceptibility from ~\cite{Miura} is show in (a) by solid line. Dashed lines are guides to the eye. Vertical lines indicate the first mode maximum temperature and the proposed structural transition temperature.}
\end{figure}

The temperature dependence of the peak intensity of two Lorentzians describing the dynamic susceptibility is presented in Fig.~\ref{peaks}a. The obtained temperature dependence of narrow Lorentzian height (green line in inset of Fig.~\ref{Dyn}) has a clear maximum at 525 K. Note that these measurements fixed the presence of peaks of the low-temperature phase at 550 K, which suggests a structural transition temperature between 550 and 575 K (marked by a vertical line on Fig.~\ref{peaks}). A completely different behavior is demonstrated by the wide Lorentzian. Its intensity reaches a maximum near the temperature of the structural transition and remains high in the high-temperature phase. 

The obtained relaxation profiles of both peaks of the dynamic susceptibility make it possible to estimate the relaxation times of these modes, which are inverse proportional to the half-widths of the fitted peaks at half maximum. The relaxation time of the narrow mode is an order of magnitude longer and has a pronounced maximum near 525 K (Fig.~\ref{peaks}b), as the intensity of this peak. The shorter relaxation time of a wide mode increases with increasing temperature. The fitting at the highest temperatures is less reliable due to the small amplitude of the narrow peak, which can lead to errors in determining relaxation times at these temperatures.

The appearance of low-frequency quasielastic scattering (central peak) in the phase transition temperature region is a common feature of inelastic light scattering spectra observed in various materials~\cite{Lev}. It is believed that this peak reflects some internal relaxation mode. Various dynamical mechanisms, such as entropy and phonon density fluctuations, overdamped soft modes, degenerate electronic states, molecular orientations, phasons, ion motion, and tunneling, can lead to the appearance of central peaks in the scattering of light~\cite{Lev}. 

In the spin systems quasielastic scattering has been frequently reported in Raman scattering measurements~\cite{Lem}. Usually its origin is explained in terms of either spin diffusion or spin-energy fluctuations.The latter mechanism was proposed to explain critical scattering in various antiferromagnets and was used to describe quasielastic scattering in both three-dimensional and low-dimensional materials. It has been shown by Reiter~\cite{Reit} that fluctuations in the total magnetic energy in a magnetic insulator can scatter light, leading to a peak at zero frequency. Its width is determined by the spin-lattice relaxation time, and the integrated intensity is proportional to the magnetic contribution to the heat capacity. The mechanism describing such a contribution to the scattering cross section also determines two-magnon light scattering, i.e. it is associated with scattering by pairs of spin fluctuations. According to the theory of Reiter~\cite{Reit} and Halley~\cite{Hal}, dynamic susceptibility can be expressed as:
\begin{equation}
\label{dyn-chi}
    \chi''(\omega)/\omega\propto C_m T\frac{\gamma+D_Tq^2}{\omega^2+(\gamma+D_Tq^2)^2}
\end{equation}
where $C_m$ is the magnetic specific heat, $D_T$ is the thermal diffusion constant, $q$ is wave vector and $\gamma$ is spin-lattice relaxation rate. The term $D_Tq^2$ is added if the scattering does not occur exactly for $q = 0$ to account for diffusion. Expression \eqref{dyn-chi} represents Lorentzian with a half-width $\gamma+D_Tй^2$. It is this form of two lines of quasielastic scattering with $\omega$ = 0 that was obtained by fitting the dynamic susceptibility in  Li$_2$RuO$_3$. Due to the lack of experimental capabilities, it is currently not possible to assume the shape of the third very narrow central peak, the existence of which follows from Fig.~\ref{RSnew}.

The quasielastic scattering in broad temperature range above $T_N$ has been observed in some magnets such as a two-dimensional antiferromagnet FePS$_3$~\cite{Sec} and later  1D antiferromagnets KCuF$_3$~\cite{Onda} and CuGeO$_3$~\cite{Loo,Kur}. Since the band gap in Li$_2$RuO$_3$ is at least $\sim$0.2 eV ~\cite{Park} even in the high-temperature phase, the observed quasielastic scattering of light cannot be associated with charge electronic excitations. Also, moderate softening of the phonon modes does not imply an explanation of this scattering in the concept of a soft mode ~\cite{Ginz}. In tetragonal crystals, light scattering by fluctuations of magnetic energy is not visible in a configuration where the scattered light is polarized perpendicular to the incident polarization~\cite{Reit}. It dominates parallel polarizations, but it can be observed in crossed polarization geometry in crystals with lower symmetry. As can be seen, in our case, temperature-dependent quasielastic scattering of $A_g$ (XX, YY) symmetries dominates with some impurity of $B_g$ (XY), and both symmetries are associated with directions in the  plane of Ru hexagons.

\begin{figure}[b]
\includegraphics[width=0.5\textwidth]{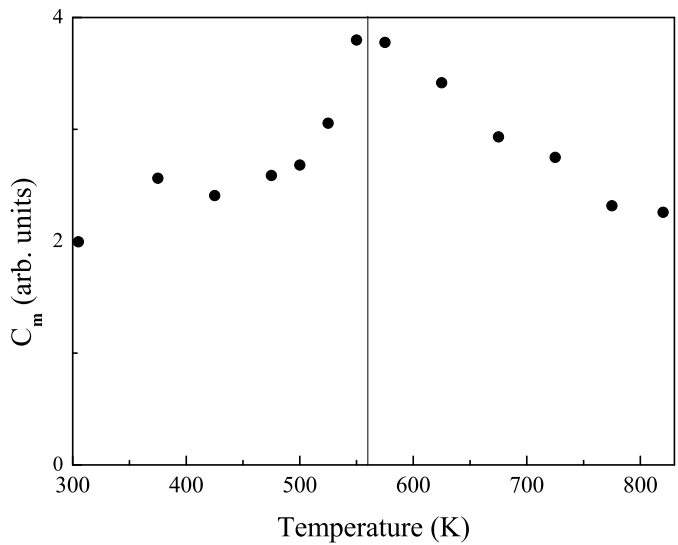}
\caption{\label{Cm}  Temperature dependence of magnetic specific heat $C_m$ derived from the peak height of the fitted Lorentzian
line for the broad feature of Li$_2$RuO$_3$ dynamic susceptibility.}
\end{figure}

The temperature behavior of the wide component of quasi-elastic light scattering correlates with the behavior of magnetic susceptibility, which suggests that it is related to the magnetic fluctuations. Using Eq. \eqref{dyn-chi} we  estimated the magnetic specific heat $C_m$ from the peak height of this component. Interestingly, it has a maximum near 560 K, i.e., the temperature of the structural transition, see Fig. ~\ref{Cm}.

If we use Eq. \eqref{dyn-chi} to describe a narrow peak, then we get a feature in magnetic part of the specific heat at a lower temperature ~ 525 K.  However, at the temperature of formation of a narrow peak, the absorption of light leads to both disorientation of the spins forming the singlet and disordering of the superstructure of the singlet. In this case, the appearance of magnetization is possible and the use of Eq. \eqref{dyn-chi} is not entirely correct. It is noteworthy that the maximum of the narrow Lorentz mode is observed on the low-temperature side of the phase transition at a temperature of $\sim$525 K. Thus, in contrast to the second relaxation mode (wide Lorentz), which retains considerable strength in the high-temperature phase, this one is associated exclusively with strong fluctuations that occur in the  P2$_1$/m phase. Their    origin is  of   course unknown,  but  one   might    speculate   that   they  are connected   with   spin   degrees    freedom  or  magneto-elastic    coupling.   Indeed, pronounced    changes   at  the transition are    observed    mostly  in    magnetic characteristics   and  are    related  to    formation  of  spin   singlets.  This fact suggests that the narrow Lorentzian of quasielastic scattering is    most    probably  due   to fluctuations in magnetic energy near the temperature of the formation of a superstructure of antiferromagnetic dimers, accompanied by a collapse of the magnetic susceptibility. In favor of such an interpretation, the observation of a similar narrow peak of the Raman susceptibility at pressures of the transition to the dimerized nonmagnetic phase in honeycomb $\alpha$-RuCl$_3$~\cite{PRM} can serve. However, in the low-temperature phase, magnetization fluctuations can also contribute to scattering, leading to a modification of Eq. \eqref{dyn-chi}~\cite{Reit,Hal,Onda}. It is possible that the decay times for the energy and magnetization are different, as well as the decay times for spin-lattice and diffusion scattering mechanisms. This may explain the reason for the appearance of two (and possibly three) modes with different widths.

One example of the mechanism of the appearance of a central peak in Raman scattering due to the removal of the degeneracy of electronic levels in the presence of a spin-phonon interaction is terbium vanadate TbVO$_4$~\cite{Tb}, where the energy costs of lattice distortion are overcompensated by a decrease in electron energy in the low symmetry phase. Perhaps the confirmation of this scenario is the appearance of high-frequency electronic excitation at T $\leq$ 525K (as we will see later).

Thus, although the ground state of Li$_2$RuO$_3$ is not a classical antiferromagnet and its spectrum of magnetic excitations is unknown, it can be assumed that the origin of the complex spectrum of quasi-elastic scattering, consisting of 3 components, is due to fluctuations in magnetic energy. The results suggest that a true magnetic transition, determined by the formation of the superstructure of spin singlets, occurs at a temperature slightly lower than the structural transition, which is consistent with the conclusions of ~\cite{Mech}.

\subsection{High-frequency electronic Raman spectra of Li$_2$RuO$_3$}

\begin{figure}[t]
\includegraphics[width=0.40\textwidth]{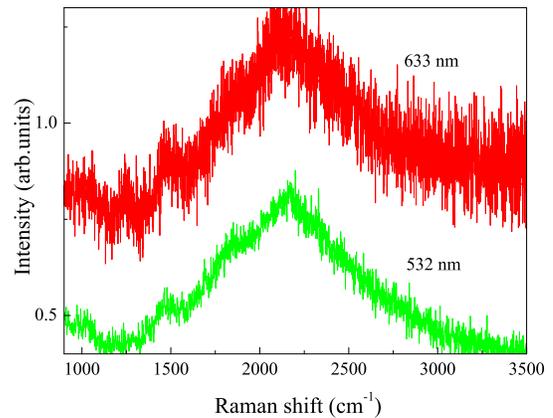}
\caption{\label{laser} Room temperature spectra of high-frequency Raman peak at 2150 $cm^{-1}$ measured with different excitation energies.}
\end{figure}

\begin{figure}[t]
\includegraphics[width=0.40\textwidth]{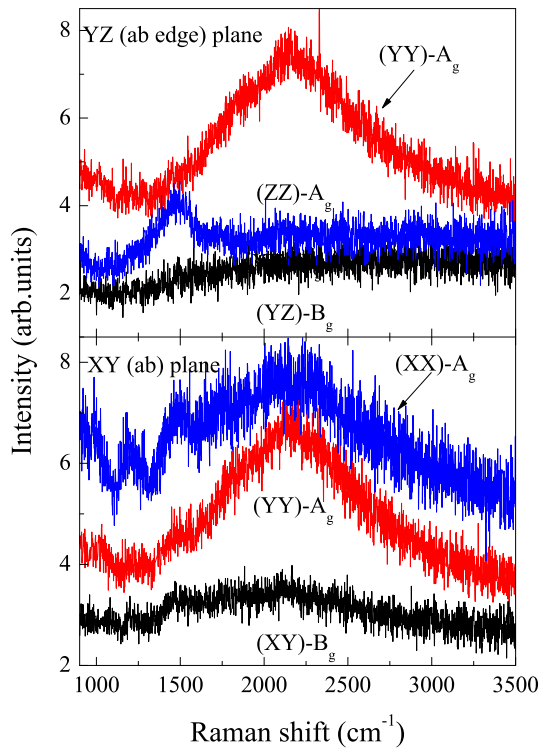}
\caption{\label{symm} Room temperature spectra of high-frequency Raman peak at 2150 $cm^{-1}$ measured in different polarization geometries.}
\end{figure}

In addition to the phonon lines located in the spectral region up to 700 $cm^{-1}$, we found a broad band of inelastic scattering of light at 2150 $cm^{-1}$, which is very similar to the previously observed scattering in SrRu$_2$O$_6$\cite{Sr19}. In order to rule out luminescence as the origin of the high-frequency broad band, Raman spectra were recorded with a different laser lines (633 and 532 nm). Observation of this peak with excitation by different lasers (Fig.~\ref{laser}) confirms that this peak is an electronic Raman scattering. Polarization measurements showed that the strong signal is observed in the A$_g$ symmetry (YY, XX). Much less intensity is found in the B$_g$ symmetry (XY) and it is absent in the geometries of the ZZ (out-of-plane A$_{g}$ and ZY B$_g$ (Fig.~\ref{symm}). This indicates that only components of the scattering tensors in the honeycomb plane have nonzero values.  A rather weak bands were observed near 1200 and 1450 cm$^{-1}$, which have frequency twice as large as group of  Raman peaks near 600-700 $cm^{-1}$. Like the lines of the one-phonon spectrum, they soften and substantially broaden with increasing temperature. We believe that these lines as in SrRu$_2$O$_6$ are due to two-phonon Raman scattering.

Thus, within the $C_{2h}$ point group, the dominating symmetry of 2150 $cm^{-1}$ peak is $A_{g}$ with a little admixture of the $B_g$. All symmetry allowed transitions  in this case are the following: $A_{g} \leftrightarrow A_{g}$, $B_{g} \leftrightarrow B_{g}$ and $B_g \leftrightarrow A_g$. The available calculations of the electronic structure of Li$_2$RuO$_3$\cite{Mazin} suggest that these transitions can occur between the  Ru $t_{2g}$ orbitals located below and above the Fermi level. According to this calculation the $xy$ orbitals on two Ru ions forming a dimer are directed, in this edge-sharing geometry, exactly toward each other, which results in the strong bonding-antibonding splitting ($\sim$ 2 eV). In contrast, the overlap between $xz/yz$ orbitals is not that large, and hence, the splitting for these orbitals is small, and they behave mostly as site-centered atomic orbitals. Available DFT results\cite{Miura,Mazin} indeed show that one might expect low-energy $d-d$ excitations with energies $\sim$200-300 meV.

\begin{figure}[b]
\includegraphics[width=0.45\textwidth]{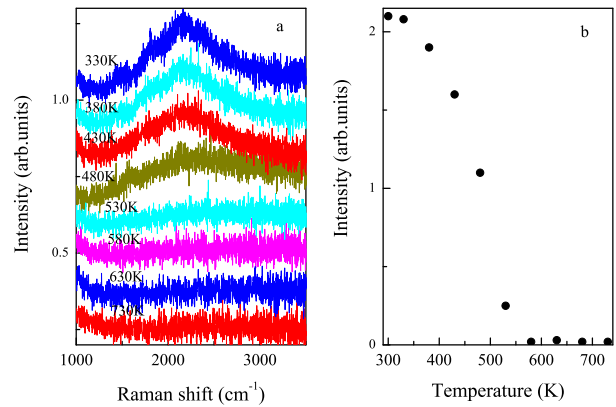}
\caption{\label{elsc_T} Spectra of high-frequency Raman peak at 2150 $cm^{-1}$ at different temperatures (a) and the  temperature dependence of its intensity (b).}
\end{figure}

The temperature evolution of the peak at 2150 $cm^{-1}$ is shown in Fig.~\ref{elsc_T}. Its intensity decreases significantly when approaching the temperature of the  transition and it is not observed at T$\geq$T$_C$. The decrease in peak intensity correlates with the corresponding decrease in the difference in Ru-Ru distances with the temperature increase~\cite{Mazin}. In addition, the peak appears and begins to grow in intensity with decreasing temperature just in the temperature region, where a narrow relaxation mode develops (Fig.~\ref{peaks}) supposedly associated with fluctuations during the formation of the spin singlet superstructure. Moreover, the symmetry of quasielastic light scattering corresponds to the symmetry of high-frequency scattering. The appearance of high-frequency excitation due to d-d transitions and its correlation with a narrow relaxation mode confirms the idea that it is an orbital degeneracy induces spontaneous dimerization of spins in Li$_2$RuO$_3$ by the formation of the Ru-Ru molecular orbitals ~\cite {Jackeli}. On the other hand, the absence of a peak at T$\geq$T$_c$ is not consistent with the possible existence of dimer liquid in the high-temperature phase ~\cite{Mazin}.

\section{Conclusion} 

The nonphononic inelastic scattering of light in Li$_2$RuO$_3$ manifests itself through both the presence of a quasielastic continuum in the phonon region of the spectrum and the formation of a high-frequency band near 2150  $cm^{-1}$. The temperature behavior of both phenomena is rather complicated. Two peaks in dynamic susceptibility demonstrate a significant difference in the temperature behavior of both their intensities and relaxation times. The first one having a maximum near 525 K is associated with formation of the superstructure of spin singlets. The second one, which has a relaxation time an order of magnitude shorter, retains considerable intensity at T$\geq$T$_C$; its fitting by the mechanism of the magnetic energy fluctuation gives the maximum in magnetic heat capacity near the structural transition temperature of 560 K. 

The intensity of the high-frequency peak near 2150  $cm^{-1}$, on the contrary, decreases and it is not observed at T$\geq$T$_C$. Most probably, this peak is due transitions between $4d$ orbitals of Ru, the degeneracy of which is removed in the low-temperature phase during the formation of Ru spin singlets. Since the obtained data suggest the presence of a third relaxation component with a long relaxation time, additional studies of the critical dynamics Li$_2$RuO$_3$ are necessary, as well as further theoretical analysis of these results.
 
\section{Acknowledgements}
The research was carried out  within the state assignment of Ministry of Science and Higher Education of the Russian Federation (No. AAAA-A18-118020190098-5, topic "Electron" and No. AAAA-A18-118020190095-4, topic “Quantum”) and was partially supported by the grants of the Russian Foundation for Basic Research (projects no. 19-52-18008 and 20-32-70019).

\end{document}